\documentstyle[twocolumn,aps,epsf]{revtex} 
\begin{document}
\def\abstract#1{\begin{center}{\large Abstract}\end{center} \par #1}
\newenvironment{proof}{\noindent{\em Proof.}}%
{{\hspace*{1em}\hfill{$\Box$}}}
\title{\bf String excitation inside generic black holes} 
\author{Kengo Maeda \thanks{Electronic address: 
g\_maeda@gravity.phys.waseda.ac.jp}, 
Takashi Torii \thanks{Electronic address: torii@th.phys.titech.ac.jp}, 
Makoto Narita \thanks{Electronic address: narita@se.rikkyo.ac.jp}} 
\address{${}^{\mbox{\rm *}}$
Department of Physics, Waseda University, Shinjuku-ku, Tokyo 169, Japan}
\address{${}^{\mbox{\rm \dag}}$
Department of Physics, Tokyo Institute of Technology, Oh-Okayama, 
Meguro-ku, Tokyo 152, Japan}
\address{${}^{\mbox{\rm \ddag}}$
Department of Physics, Rikkyo University, Nishi-Ikebukuro, Toshima-ku, 
Tokyo 171, Japan}
\maketitle 
\abstract{We calculate how much a first-quantized string is excited 
after crossing the inner horizon of charged Vaidya solutions, 
as a simple model of generic black holes. 
To quantize a string suitably, we first show that the metric is approximated 
by a {\it plane-wave} metric near the inner horizon when the surface 
gravity of the horizon $\kappa_I$ is small enough. 
Next, it is analytically shown that the string crossing the inner horizon is 
excited infinitely in an asymptotically flat spacetime, while it is finite
in an asymptotically de Sitter spacetime and the string can pass across 
the inner horizon when $\kappa_I<2\kappa:= 
2\,\mbox{min}\{\kappa_B,\kappa_C \}$, where $\kappa_B$~($\kappa_C$) is the 
surface gravity of the black hole~(cosmological) event horizon. 
This implies that the strong cosmic censorship holds in an asymptotically 
flat spacetime, while it is violated in an asymptotically de Sitter 
spacetime from the point of view of string theory.}\\  
\section{Introduction}
One of the interesting issues in general relativity is the inner structure 
of {\it generic} black holes, related to the strong cosmic censorship 
conjecture~\cite{Penrose}. The conjecture states that every physically 
reasonable spacetime is globally hyperbolic, or it is uniquely determined 
by initial regular data on a spacelike hypersurface $S$. Therefore, if 
the conjecture is violated, the spacetime has a Cauchy horizon~(CH), which is 
the boundary of the future~(past) domain of dependence of $S$ and we cannot 
predict what happens for an observer crossing the CH. 

Over the past few years a considerable number of studies have been made 
on the internal structure of generic charged or rotating black holes 
both analytically and numerically. 
Poisson and Israel~(PI)~\cite{PoissonIsrael} showed that a scalar curvature 
singularity appears generically instead of the CH by using a simple 
model of spherically symmetric charged black holes. 
Ori~\cite{Ori} constructed an exact solution of the Einstein-Maxwell equations 
for the model, which suggests that the CH is transformed generically into a 
null weak singularity. 
This was verified numerically in charged black holes~\cite{BradySmith,Burko}. 
Brady and Chambers~\cite{BradyChambers} extended these arguments to the 
case of more realistic black hole models and suggested that 
the CH for Kerr-type vacuum black 
holes is transformed generically into a null weak singularity, where 
the local geometry and the strength of the singularity are quite similar 
to the spherically symmetric charged black holes. Thus, this series of works 
strongly suggests that the strong cosmic censorship holds in the framework of 
general relativity in the sense that there is no regular CH inside generic 
black holes. However, there remain unsettled questions. 
Is such a null weak singularity a real singularity, or the end of spacetime 
in a quantum theory of gravity? In particular, we have no precise 
knowledge about what happens from the point of view of quantum gravity 
in such a region where the curvature is very strong. 

One of the promising candidates for a quantum theory of gravity is string 
theory. 
Horowitz and Steif~\cite{HS1,HS2} have proposed a new criterion for a 
singularity in terms of a first-quantized test string, namely if the 
expectation value of the mass associated with the test string diverges 
at a finite time, then spacetime is called singular. 
This is an extension of the classical definition of 
singularity~\cite{HE}. As already shown in~\cite{HS1,Guven,AmatiKlimcik},
all solutions to the vacuum Einstein equation with a covariantly constant 
null vector are also solutions to the classical equations of motion for 
the metric in string theory. These solutions are known as 
{\it plane-waves}~\cite{HS1}. 
Thus, if the metric for Kerr-type generic vacuum black holes is 
approximated by a plane-wave metric near the ``singular CH''~(hereafter, 
simply called CH), it is also the metric for a classical solution 
of string theory near the CH. 
In this case, it is worth testing whether such a null weak 
singularity is also a singularity for the first-quantized string, as a first 
step for considering the quantum corrections. 

In this paper, we test how much the first-quantized string gets excited 
after it crosses the CH in spherically symmetric charged black holes 
instead of Kerr-type generic ones for simplicity. 
As mentioned above, the spherically symmetric charged black hole is a 
good and simple model for describing the internal structure of the 
Kerr-type black hole because their internal structures are quite similar to 
each other. 
Firstly, we show that the spacetime near the CH is described approximately 
by a plane-wave metric if the surface gravity $\kappa_I$ of the inner 
horizon is small enough~(section~III). 
Secondly, we calculate explicitly the expectation value of the mass of 
the test string in two cases: (i) the charged black hole is embedded in 
flat spacetime; (ii) it is embedded in de Sitter 
spacetime~(section~IV). 
Finally, we discuss the strong cosmic censorship in the framework of 
string theory on the basis of the previous results~(section~V). 
In the following section, we start to review briefly the PI model 
of spherically symmetric charged black holes. 

\section{Null weak singularities along the Cauchy horizon}
When a physically realistic gravitational collapse with charge $q$ and/or 
angular momentum $a$ occurs, gravitational and/or electromagnetic 
waves propagate outward and some of them are back-scattered by the 
effective potential of the gravitational field outside a black hole. 
This implies that there exists, at least, two types of null flux near the CH; 
an ingoing null flux along the CH and an outgoing null flux produced by 
the scattering of the ingoing null flux inside the black hole. 

PI~\cite{PoissonIsrael} constructed a charged spherically symmetric model 
which well describes the interior of generic black holes with an inner 
horizon. They showed that once the CH is contracted 
by the outgoing flux, the invariantly defined quasi-local mass diverges and 
a curvature singularity inevitably appears. 
This is the consequence of the nonlinear interaction between the outgoing 
and the infinitely blue-shifted ingoing null fluxes. The outgoing 
null flux, however, is not important, as firstly pointed out by PI. 

Before introducing the PI model, let us investigate a charged Vaidya solution, 
where only a purely-ingoing null flux exists. The spacetime is described by 
the metric 
\begin{eqnarray}
\label{eq-Vaidya}
ds^2 = d\omega \,(2dr -f d\omega)+r^2(d\theta^2+ \sin^2 \theta d\phi^2), 
\end{eqnarray}
\begin{eqnarray}
\label{eq-Vaidyaf}
f = 1-\frac{2m(\omega)}{r}+\frac{q^2}{r^2}-\frac{\Lambda}{3}r^2, 
\end{eqnarray}
where $q$ is an electric charge and $\Lambda$ is the positive 
cosmological constant. The energy-momentum tensor is 
\begin{eqnarray}
\label{eq-energy}
T_{\mu\nu}&=&\rho_{\rm{in}}\,l_\mu l_\nu + E_{\mu\nu}, 
\end{eqnarray}
\begin{eqnarray}
\label{eq-energyE}
E_{\mu\nu}&=&2F_{\mu\alpha}{F_\nu}^{\alpha}-\frac{1}{2}g_{\mu\nu}F^2, 
\end{eqnarray}
where $\rho_{\rm{in}}$ represents the energy density of the ingoing null flux 
along the ingoing radial null vector, $l^\mu\equiv (\partial_r)^\mu$. 
Here, $F_{\mu\nu}$ is a purely electric Maxwell field 
$F=(q/r^2)\,d\omega\wedge dr$. 
From the conservation law, $\rho_{\rm{in}}$ is simply related to the 
mass $m(\omega)$ as follows, 
\begin{eqnarray}
\label{eq-density}
\rho_{\rm{in}}=\frac{1}{4\pi r^2}\frac{dm}{d\omega}. 
\end{eqnarray}
Now, we consider a freely-falling observer into the inner horizon 
along radial timelike geodesics whose tangent vector is 
$u=(\dot{r},\dot{w},0,0)$, where a dot denotes the derivative with 
respect to the proper time $\tau$. 
The relation between $\omega$ and $\tau$ can be obtained by the following 
geodesic equation 
\begin{eqnarray}
\label{eq-omega-tau}
2\,\ddot{\omega}=-f_{,r}\,{\dot{\omega}}^2. 
\end{eqnarray}
\begin{figure}[htbp]
 \centerline{\epsfxsize=6.0cm \epsfbox{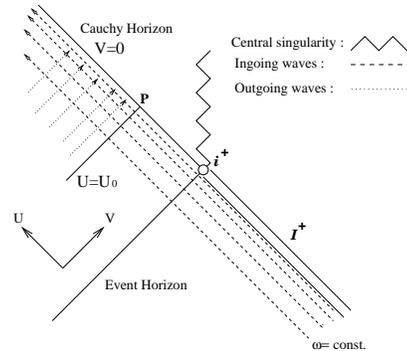}}
         \caption{The Penrose diagram of a spherically symmetric charged 
black hole with outgoing and infinitely blue-shifted ingoing null fluxes 
for $U\ge U_0$. A null weak singularity appears along $V=0$.}
               \protect
\label{fig-cauchy}
\end{figure}
Define the inner horizon $r=r_I$ and the surface gravity $\kappa_I$ as 
\begin{eqnarray}
\label{eq-indef}
f(r_I,m_0) = 0, \quad 
\kappa_I := -\frac{1}{2} f_{,r}(r_I, m_0), 
\end{eqnarray}
where $m_0$ is the asymptotic Bondi-like 
mass,~i.e.~$m_0:= m(\infty)$. It must be noted that the inner 
horizon corresponds to the CH in the Vaidya metric~(\ref{eq-Vaidya}). 
By solving Eq.~(\ref{eq-omega-tau}), $\dot{\omega}$ behaves as 
\begin{eqnarray}
\label{eq-omega}
\dot{\omega} \sim -\frac{1}{\kappa_I \tau} 
\end{eqnarray}
near the CH. From Eqs.~(\ref{eq-energy}) and (\ref{eq-density}) 
the energy density $\rho_{\rm{ob}}$ seen by the freely-falling 
observer is given by 
\begin{eqnarray}
\label{eq-ob-energy}
\rho_{\rm{ob}} = T_{\tau \tau} \sim T_{\omega\omega}\,{\dot\omega}^2 \sim 
\frac{1}{4\pi (\kappa_I{r_I} \tau)^2 }\frac{dm}{d\omega}. 
\end{eqnarray}

Next, to observe the non-linear interaction between the outgoing and ingoing 
null fluxes, we introduce double null coordinates 
\begin{eqnarray}
\label{eq-double}
ds^2 = -2e^{-\lambda(U,V)}dU dV + {r(U,V)}^2 (d\theta^2+\sin^2{\theta} d\phi^2) \end{eqnarray}
near the point $P$~(See Fig.~\ref{fig-cauchy}), where 
$\partial_U$ and $\partial_V$ are ingoing and outgoing null vectors, 
respectively. The quasi-local mass $M$ is defined by 
\begin{eqnarray}
\label{eq-mass}
1-\frac{2M(U,V)}{r}+\frac{q^2}{r^2}-\frac{\Lambda}{3}r^2 
:= g^{\mu\nu}r_{,\mu}r_{,\nu}. 
\end{eqnarray}
Defining outgoing and ingoing expansions along outgoing and 
ingoing null geodesics by 
$\theta_{-}:=2r_{,U}/r$ and $\theta_{+}:= 2r_{,V}/r$, 
respectively, the asymptotic behavior of $M$ near the CH~($V=0$) is 
\begin{eqnarray}
\label{eq-mas}
M \sim \theta_{-}\theta_{+} + (\mbox{finite terms}).  
\end{eqnarray}
The Raychaudhuri equation~\cite{HE} along $U=U_0$ becomes  
\begin{eqnarray}
\label{eq-Rayv}
\frac{d\theta_{+}}{dV} = -\frac{1}{2}{\theta_+}^2-
T(\partial_V,\partial_V)
\end{eqnarray}
where we choose $V$ such that $\lambda=\mbox{const}.$ along $U=U_0$. 
Because $\tau\sim V$ and $T(\partial_V,\partial_V)
\sim \rho_{\rm ob}$, $M$ near the 
CH can be estimated roughly as 
\begin{eqnarray}
\label{eq-masrho}
M \sim \theta_{-}\int\rho_{\rm ob}d\tau + (\mbox{finite terms})  
\end{eqnarray}
through Eq.~(\ref{eq-Rayv}). 
Let us suppose that there exists a positive outgoing null flux 
$L(U):= T(\partial_U,\partial_U)>0$ crossing the CH after $U\ge U_0$ 
and reparametrize the coordinate $U$ such that $U$ is the affine 
parameter of the null geodesic generator of the CH. 
Then, through the Raychaudhuri equation along the CH 
\begin{eqnarray}
\label{eq-Rayu}
\frac{d\theta_{-}}{dU} = -\frac{1}{2}{\theta_-}^2-L(U), 
\end{eqnarray}
$\theta_{-}$ becomes negative after the outgoing flux passes 
through the CH because $\theta_{-}=0$ before $U<U_0$. 
This implies that $M$ diverges when the integral of $\rho_{\rm{ob}}$ 
with respect to $\tau$ becomes infinite. 
This is called {\it mass inflation}. 
The mass inflation corresponds to the scalar polynomial curvature 
singularity~({\it s.p. curvature singularity}) because the square of 
the Weyl curvature behaves as 
$C_{\alpha\beta\gamma\delta}C^{\alpha\beta\gamma\delta}\sim M^2$. 

As shown in Ref.~\cite{Ori}, this null singularity is a weak singularity 
in the sense that no null object falling into the singularity can be 
crushed to zero area, namely the $2$-$d$ area of the CH is non-zero. 
Here, we should not overlook the following facts; (i) the outgoing null flux 
can only contract the CH slowly, and the infinitely blue-shifted 
ingoing null flux is essential for the singular behavior along the CH, 
(ii) the Vaidya solution~(\ref{eq-Vaidya}) contains a 
{\it p.p. curvature singularity}~(any of the components of 
$R_{abcd}$ diverges along a 
parallel propagated basis)~\cite{HE} at the CH whenever $\rho_{\rm ob}$ 
diverges, although no s.p. curvature singularity appears in the 
metric. 
This implies that the Vaidya metric is a good approximation for representing 
the singular behavior along the CH near the timelike infinity $i^+$ 
where $\theta_{-}$ at the CH is negligibly small. 

Hereafter, we shall consider the propagation of a test string on the 
Vaidya metric. In the next section, we will show that the Vaidya 
metric~(\ref{eq-Vaidya}) is approximated by a plane-wave metric near 
the CH if $\kappa_I$ is small enough. 

\section{Plane symmetric approximation} 

Firstly, we shall consider the transformation from the 
Vaidya metric~(\ref{eq-Vaidya}) to the double null 
metric~(\ref{eq-double}). 
Let us define a null coordinate $u$ such that  
\begin{eqnarray}
\label{eq-u-co}
du=\frac{2g}{f}dr -g\,d\omega, 
\end{eqnarray}
where $g$ is a function of $r$ and $\omega$. From the coordinate 
condition $d^2 u=0$, 
\begin{eqnarray}
\label{eq-u-cond}
2\dot{G} + f G'=-f' G, 
\end{eqnarray}
where $G$ is defined by $g=fG$. A dot and a prime denote the 
derivatives with respect to $\omega$ and $r$, respectively. 
This is a linear partial differential equation for $G$. Following the 
standard procedure, let us consider the characteristic curve 
obeying the equation 
\begin{eqnarray}
\label{eq-character}
\frac{d\omega}{2}=\frac{dr}{f}=-\frac{dG}{f'G}. 
\end{eqnarray}
Hereafter, we consider just the neighborhood of the CH,~i.e., 
$\omega\gg 1 \gg y:=r-r_I$. Using 
$f\simeq -2\kappa_I\,y -2(m_0-m(\omega))/r_I$, the first equality can be 
reduced to the following ordinary differential equation near the CH, 
\begin{eqnarray}
\label{eq-G1-or}
\frac{dy}{d\omega} + \kappa_I y \simeq \frac{\delta m(\omega)}{r_I}, 
\end{eqnarray}
where $\delta m(\omega):= m_0-m(\omega)$. 
The solution is 
\begin{eqnarray}
\label{eq-G1-so}
e^{\kappa_I\omega}\,y - \int_{\omega_0}^{\omega} 
\frac{\delta m(\omega')}{r_I} e^{\kappa_I\omega'}d\omega'\simeq\mbox{const}. 
\end{eqnarray}
By solving the other equation~(\ref{eq-character}), 
$G$ approximately behaves as 
\begin{eqnarray}
\label{eq-G2-so}
G e^{-\kappa_I \omega} \simeq \mbox{const}. 
\end{eqnarray}
near the CH. 
Combining the above two solutions, the general form of the solution of 
Eq.~(\ref{eq-u-cond}) is 
\begin{eqnarray}
\label{eq-G-general}
G\simeq e^{\kappa_I \omega} F[X], 
\end{eqnarray}
\begin{eqnarray}
\label{eq-G-generalX}
X := e^{\kappa_I\omega} y - 
\int_{\omega_o}^\omega \frac{\delta m(\omega')}{r_I}
e^{\kappa_I\omega'}d\omega' \nonumber 
\end{eqnarray}
near the CH. By taking $g=1$ for $\delta m(\omega)=0$, 
$F[X]$ can be fixed as 
\begin{eqnarray}
\label{eq-Ffix}
F[X]=-\frac{1}{2\kappa_I X}. 
\end{eqnarray}

Now, let us determine the coordinate $u$. From Eq.~(\ref{eq-u-co}), 
$u(r,\omega)$ satisfies the following equations, 
\begin{eqnarray}
\label{eq-u-con1}
\frac{\partial u}{\partial r}=2G, \qquad 
\frac{\partial u}{\partial \omega}=-f G 
\end{eqnarray}
and hence 
\begin{eqnarray}
\label{eq-u-con2}
f \frac{\partial u}{\partial r} + 2\frac{\partial u}{\partial \omega}= 0. 
\end{eqnarray}
Applying the previous procedure to Eq.~(\ref{eq-u-con2}), the general 
form of the solution is 
\begin{eqnarray}
\label{eq-u-form}
u \simeq H[X] 
\end{eqnarray}
near the CH, where $H[X]$ is an arbitrary function of $X$. To determine 
the function $H[X]$, we use the first of Eqs.~(\ref{eq-u-con1}) 
and find that 
\begin{eqnarray}
\label{eq-H-solu}
H[X]=-\frac{1}{\kappa_I}\ln |X|. 
\end{eqnarray}
Then, $u$ behaves as 
\begin{eqnarray}
\label{eq-u-solu}
u \simeq -\omega-\frac{1}{\kappa_I}\ln \left| y-e^{-\kappa_I \omega} 
\int_{\omega_0}^{\omega}\frac{\delta m(\omega')}{r_I}e^{\kappa_I\omega'}
d\omega' \right| 
\end{eqnarray}
and the Vaidya metric~(\ref{eq-Vaidya}) can be rewritten in terms 
of the new coordinate $u$ as 
\begin{eqnarray}
\label{eq-u-solu1}
ds^2\simeq \frac{1}{G} du\, 
d\omega + (y+r_I)^2 (d\theta^2 + \sin^2\theta d\phi^2). 
\end{eqnarray}

Secondly, we introduce Kruskal like coordinates $U,\,V$ such as 
\begin{eqnarray}
\label{eq-UVkraskal}
U=-e^{-\kappa_I u}, \qquad V=-e^{-\kappa_I \omega}. 
\end{eqnarray}
By Eqs.~(\ref{eq-G-general}) and~(\ref{eq-u-solu}), 
the metric~(\ref{eq-u-solu1}) is reduced to the double null coordinates 
\begin{eqnarray}
\label{eq-UVcoods}
ds^2\simeq -\frac{2}{\kappa_I}dU dV + (y+r_I)^2 
(d\theta^2 + \sin^2\theta d\phi^2), 
\end{eqnarray}
where 
\begin{eqnarray}
\label{eq-yUV}
y\simeq UV -\frac{V}{\kappa_I r_I}\int_{V_0}^{V}
\delta m{\left(-\frac{\ln(-V)}{\kappa_I} \right)}\frac{dV}{V^2}. 
\end{eqnarray}

Finally, we bring the above metric into the plane-wave form, 
following the same procedure as in Ref.~\cite{HorowitzRoss}. 
If $r_I$ is large enough compared with the typical size of 
a test string crossing the CH, 
we can approximate the two sphere metric 
$(y+r_I)^2 d\Omega$ by a flat metric such as 
\begin{eqnarray}
\label{eq-ytilde{U}V}
ds^2 \approx -d\tilde{U}dV + (y+r_I)^2 dX_{i}dX^{i}, 
\end{eqnarray}
where $\tilde{U}=2 U/\kappa_I$ and $X^i\,(i=1,2)$ are transverse 
Cartesian-like coordinates. As it can be easily verified, 
$\nabla_\mu l_\nu \sim O(\kappa_I)$, where $l_\mu\equiv \partial_\mu V$. 
This implies that when the surface gravity $\kappa_I$ is small enough, 
the first term of Eq.~(\ref{eq-yUV}) is negligible and we can take 
the light cone gauge~(for more detail, see Ref.~\cite{HS2}). 
Thus, for simplicity, we shall consider the $\kappa_I\ll 1$ case and 
ignore the first term of Eq.~(\ref{eq-yUV}). 
Following the methods of Ref.~\cite{Gibbons}, we transform the 
coordinates $(\tilde{U},V,X^i)$ to $(u,v,x^i)$ as 
\begin{eqnarray}
\label{eq-trans}
v &=& V, \nonumber \\
u &=& \tilde{U} + {y'}\,(y+r_I)\, X_i X^i, \\
x^i &=& (y+r_I)X^i, \nonumber 
\end{eqnarray}
where a prime means the derivative with respect to $v$. 
Then, the {\it plane-wave} metric becomes  
\begin{eqnarray}
\label{eq-plane-wave}
ds^2 &\approx& -du\,dv + H(v) x_i x^i dv^2 +dx_i dx^i, 
\end{eqnarray}
\begin{eqnarray}
\label{eq-plane-waveH}
H(v) &=& \frac{y''}{y+r_I}.
\end{eqnarray}

\section{The excitation of a test string}
In this section, we will derive the equations of motion of  
a test string and see how much the first-quantized string gets excited 
after crossing the CH, or propagating through the spacetime region 
with the infinitely blue-shifted ingoing flux. 
Following the definition of singularity for the first-quantized 
string~\cite{HS2}, we may say that the CH is singularity-free 
if the string can cross the CH without being excited infinitely. 
Conversely, if it is excited infinitely by the background field, 
then we may say that the CH has a real singularity and that it is the 
boundary of the spacetime. 
It is worth noting that all timelike geodesics end at the CH 
unless there is no curvature 
singularity~(p.p. or s.p. curvature singularities) on it, 
according to the usual interpretation of singularity~\cite{HE}. 

Let us introduce coordinates $(\tau, \sigma)$ on the world sheet 
of the test string. Then, we can take the light-cone gauge $v=P\tau$, 
which has the advantage of being unitary, where $P$ is a positive constant. 
Hereafter, we will use the same approach and conventions 
as  in Ref.~\cite{HS2}. 

In the light cone gauge, the motion of $u(\tau,\sigma)$ is 
given by 
\begin{eqnarray}
\label{eq-Umotion}
P\dot{u} = (\dot{x}^i)^2 + ({x^i}')^2 + HP^2, 
\end{eqnarray}
\begin{eqnarray}
Pu' = 2\dot{x}_i {x^i}', 
\end{eqnarray}
where a dot and a prime mean the derivatives with respect to $\tau$ and 
$\sigma$, respectively. Thus, we quantize only $x^i$ because $x^i$  
is the only independent variable in the light cone gauge. 
Because we consider closed strings here, $x^i$ is decomposed as 
\begin{eqnarray}
\label{eq-decompose}
x^i(\tau,\sigma)=\sum_{n=-\infty}^{+\infty} x_n^i(\tau) e^{in\sigma} 
\end{eqnarray}
and the field equations are 
\begin{eqnarray}
\label{eq-field}
{\ddot{x}_n}^i + n^2 x_n^i -H P^2 x_n^i=0. 
\end{eqnarray}
Because $x^i$ is real, $x_{-n}^i={x_n^i}^\ast$. 
Let us denote the independent solutions of Eq.~(\ref{eq-field}) by 
$u_n^i$ and ${\tilde{u}_n}^i$ and write $x_n^i$ as 
\begin{eqnarray}
\label{eq-atildea}
x_n^i = \frac{i}{2\sqrt{n}}(a_n^i u_n^i-
\tilde{a}_n^{i \dagger} {\tilde{u}_n}^i) 
\end{eqnarray} 
for positive $n$, where $a_n^i$ and $\tilde{a}_n^{i \dagger}$ 
are some coefficients, which are naturally interpreted as 
annihilation and creation operators, respectively, satisfying the standard 
canonical commutation relations. If $H=0$, $u_n^i$ and $\tilde{u}_n^i$ 
become  
\begin{eqnarray}
\label{eq-utildeu}
u_n^i=e^{-in\tau}, \qquad \tilde{u}_n^i=e^{in\tau}, 
\end{eqnarray} 
which represent right and left oscillators, respectively. 

We shall consider a spacetime with sandwich 
waves, namely $H=0$ for $v<-v_0$ and $v\ge 0$, as in~\cite{HorowitzRoss}. 
Physically, the latter condition implies that there is no flux from any 
pathological regions~(if they exist) like timelike singularities or infinity 
when $v\ge 0$. As a first example, we estimate the excitation of the 
string in the $\Lambda=0$ case. 
Secondly, we consider the $\Lambda>0$ case and explicitly calculate 
the excited energy of the string in some parameter ranges of $m$, $q$, 
$\Lambda$. 

\subsection{$\Lambda=0$ case}
For a generic perturbation in the $\Lambda=0$ case~\cite{Price}, the decay of 
the ingoing null flux should be given by 
\begin{eqnarray}
\label{eq-decay0}
\delta m(\omega) = \frac{\alpha r_I}{(p-1)}(\kappa_I\omega)^{-(p-1)}, 
\end{eqnarray}
where $\alpha$ is a dimensionless positive constant and $p>12$ is an 
integer. By Eqs~(\ref{eq-omega}) and~(\ref{eq-ob-energy}), the 
integral of $\rho_{\rm{ob}}$ diverges but the second integral is 
finite. This implies that $\theta_{+}$ diverges but the integral 
is finite. Therefore, once the outgoing null flux crosses the CH, 
mass inflation occurs according to Eq.~(\ref{eq-masrho}) with non-zero area 
radius $r_I$, which is independent of the power $p$. 

After a simple calculation, $H(v)$ is found to be given by 
\begin{eqnarray}
\label{eq-H0}
H \simeq -\frac{\alpha}{\kappa_I\,r_I}\frac{[-\ln(-v)]^{-p}}{v^2}. 
\end{eqnarray}
To calculate the Bogoliubov coefficients explicitly, we shall connect 
the sandwich wave region $H\neq 0$ with a ``static'' region $H=0$ 
at $v=-P\epsilon~(\tau=-\epsilon)$, where $\epsilon$ is an infinitely 
small positive value. After calculating the Bogoliubov coefficients, 
we will take the $\epsilon \to 0$ limit. 
Then, the solution of Eq.~(\ref{eq-field}) can be written as 
\begin{eqnarray}
\label{eq-H0}
x_n^i = \frac{i}{2\sqrt{n}}(b_n^i v_n^i - \tilde{b}_n^{i \dagger} 
\tilde{v}_n^i), 
\end{eqnarray}
\begin{eqnarray}
\label{eq-H0v}
v_n^i = e^{-in(\tau+\epsilon)}, \qquad  \tilde{v}_n^i = 
e^{in(\tau+\epsilon)} 
\end{eqnarray}
for the $\tau\ge -\epsilon$ region, where $b_n^i$ and 
$\tilde{b}_n^{i \dagger}$ 
are annihilation and creation operators associated with the solutions, 
$v_n^i$, $\tilde{v}_n^i$, respectively. $b_n^i$ and 
$\tilde{b}_n^{i \dagger}$ are linearly related to $a_n^i$ and 
$\tilde{a}_n^{i \dagger}$ according to the Bogoliubov transformation 
\begin{eqnarray}
\label{eq-Brelation}
b_n^i = A_n^i a_n^i - {B_n^i}^{\ast} \tilde{a}_n^{i \dagger}, \quad 
\tilde{b}_n^i = \tilde{A_n^i} \tilde{a}_n^i 
- \tilde{B_n^i}^{\ast} a_n^{i \dagger}. 
\end{eqnarray}

We can see the degree of excitation of a string for the $\tau\ge -\epsilon$ 
region which is 
initially in the ground state by calculating the total number of 
modes $\langle N_{\rm{out}} \rangle$ and the total mass-squared 
$\langle {M^2}_{\rm{out}} \rangle$. 
$\langle N_{\rm{out}} \rangle$ and $\langle {M^2}_{\rm{out}} \rangle$ 
are given by 
\begin{eqnarray}
\label{eq-N}
\langle N_{\rm{out}} \rangle &:=& 
 2\sum_{i,n=1} \langle 0_{\rm{in}}|{b_n^i}^\dagger b_n^i 
|0_{\rm{in}} \rangle = 2\sum_{i,n=1} |B_n^i|^2, 
\end{eqnarray}
\begin{eqnarray}
\label{eq-M}
\langle {M^2}_{\rm out} \rangle &:=& 
 4\sum_{i,n=1}^{\infty} 
\langle 0_{\rm{in}}| n( b_n^i b_n^{i \dagger} + 
{\tilde{b}}_n^{i \dagger} {\tilde{b}}_n^i) 
|0_{\rm{in}} \rangle \nonumber \\
&=& 8\sum_{i,n=1}^{\infty} n |B_n^i|^2-8, 
\end{eqnarray}
where the last term in Eq.~(\ref{eq-M}) corresponds to the Casimir energy. 
Because the differential equation~(\ref{eq-field}) for $x_n^i$ is independent 
of $i$, we have only to consider the $i=1$ case. 

Now, let us denote $x_n^1=x_n$ by the following complex form, 
\begin{eqnarray}
\label{eq-complex}
x_n = e^{-i n\omega}, \qquad \omega=S(\tau)+\frac{i}{n}\ln A(\tau), 
\end{eqnarray}
where $A~(\ge 0),\,S$ are real functions of $\tau$. By substituting 
Eq.~(\ref{eq-complex}) into Eq.~(\ref{eq-field}), we can obtain the 
following two equations, 
\begin{eqnarray}
\label{eq-twocom1}
A \ddot{S} + 2 \dot{S}\dot{A} = 0, 
\end{eqnarray}
\begin{eqnarray}
\label{eq-twocom2}
\ddot{A}+A [n^2(1-\dot{S}^2) - H P^2 ] = 0. 
\end{eqnarray}
The first equation~(\ref{eq-twocom1}) is easily integrated as 
\begin{eqnarray}
\label{eq-SA}
\dot{S} A^2 = 1, 
\end{eqnarray}
where we used the boundary condition at 
$\tau=-\tau_0\,(:=v_0/P)$,~i.e. 
$x_n(\tau\le-\tau_0)=e^{-in\tau}$. 
Substituting the relation~(\ref{eq-SA}) into 
Eq.~(\ref{eq-twocom2}), we find the following differential equation for $A$, 
\begin{eqnarray}
\label{eq-Adiff}
\ddot{A}=n^2\left(\frac{1}{A^3} - A \right) + HP^2 A. 
\end{eqnarray}

We use the standard matching method to obtain $B_n$, which demands 
the continuity of $x_n$ and its derivative at $\tau=-\epsilon$. 
This immediately yields 
\begin{eqnarray}
\label{eq-mach}
|B_n|^2 = \frac{1}{4}\left(A_{\epsilon}-
\frac{1}{A_{\epsilon}} \right)^2 +\frac{{\dot{A}_{\epsilon}}^2}{4n^2}, 
\end{eqnarray}
where $A_{\epsilon}$ and 
$\dot{A}_{\epsilon}$ are the values of $A$ and $\dot{A}$ at 
$\tau=-\epsilon$, respectively. 
The above equation means that 
when $\lim_{\epsilon \to 0} A_{\epsilon}=0$, the Bogoliubov coefficient 
$|B_n|^2$ for each $n$ diverges and hence 
the total number of modes $\langle N_{\rm{out}} \rangle$ and 
the total mass-squared $\langle {M^2}_{\rm{out}} \rangle$ also diverge 
in the limit $\epsilon \to 0$. 
Therefore we shall consider only the case $A_{\epsilon} \sim C>0$. 

As $H$ diverges near the $\tau=-\epsilon$, the 
equation~(\ref{eq-Adiff}) is approximately 
\begin{eqnarray}
\label{eq-twocom3}
\ddot{A} \sim H P^2 A \sim H P^2 C. 
\end{eqnarray}
By integrating it once, we obtain 
\begin{eqnarray}
\label{eq-Adot}
\dot{A}_{\epsilon} \sim C \int^{-\epsilon} H(P\tau) P^2 d\tau \propto 
-\frac{[-\ln (\epsilon)]^{-p}}{\epsilon}, 
\end{eqnarray}
which diverges in the $\epsilon \to 0$ limit. This causes again 
the divergence of $|B_n|^2$ and $\langle N_{\rm{out}} \rangle$, 
$\langle {M^2}_{\rm{out}} \rangle$. Thus we find that the CH 
is singular in terms of a first-quantized string, as mentioned before. 
\subsection{$\Lambda > 0$ case}
For a generic perturbation in the $\Lambda > 0$ case~\cite{BradyMossMyers}, 
the decay of the ingoing flux should be dominated by 
\begin{eqnarray}
\label{eq-decayl}
\delta m(\omega)\sim \frac{\kappa_I^2}{2\kappa_B} 
e^{-2\kappa_B \omega} +C\,(>0)\times 
\frac{\kappa_I^2}{2\kappa_C} e^{-2\kappa_C \omega}, 
\end{eqnarray}
where $\kappa_B$ and $\kappa_C$ are the surface gravity of the 
black hole and of the cosmological event horizon, respectively. 
The first term of the r.h.s of Eq.~(\ref{eq-decayl}) 
comes from the backscattering of the outgoing fluctuations close to the 
black hole event horizon, while the second term comes from the neighborhood 
of the cosmological event horizon. 
The dominant term for Eq.~(\ref{eq-decayl}) depends on the 
values, $\kappa_B$ and $\kappa_C$, i.e. for $\kappa_B\le 
\kappa_C$ the first term is dominant, while it becomes 
subdominant for $\kappa_B> \kappa_C$. Defining 
$\kappa:=\mbox{min}\{\kappa_B,\kappa_C \}$, we can easily find that 
the integral of 
$\rho_{\rm{ob}}$ in Eq.~(\ref{eq-ob-energy}) 
diverges when 
\begin{eqnarray}
\label{eq-kappa-di}
2\kappa\le \kappa_I
\end{eqnarray}
and the integral converges but $\rho_{\rm{ob}}$ itself diverges when 
\begin{eqnarray}
\label{eq-kappa-con}
\kappa < \kappa_I < 2\kappa. 
\end{eqnarray}
It must be noted that the first inequality in Eq.~(\ref{eq-kappa-con}) 
is always satisfied because $\kappa_B < \kappa_I$ for $\kappa=\kappa_B$ 
unless $r_I=r_B$, and $\kappa_C<\kappa_B<\kappa_I$ for 
$\kappa=\kappa_C$. Therefore the Vaidya metric~(\ref{eq-Vaidya}) always 
contains a p.p. curvature singularity at the CH.  
In the former case, mass inflation occurs provided that the outgoing 
null flux exists, but in the latter case it does not. 

By the definition of Eq.~(\ref{eq-plane-waveH}), it follows that 
\begin{eqnarray}
\label{eq-H2}
H(v) \sim 
-\frac{\alpha}{{r_I}^2}(-v)^{\frac{2(\kappa-\kappa_I)}{\kappa_I}}, 
\end{eqnarray}
where $\alpha$ is a positive constant. 
As discussed in the $\Lambda=0$ case, the Bogoliubov coefficient 
$|B_n|^2$ diverges when $\kappa_I\ge 2\kappa$ and hence 
$\langle N_{\rm{out}} \rangle$, $\langle {M^2}_{\rm{out}} \rangle$ also 
diverge, which implies that the singularity at the CH is also a 
singularity in terms of a first-quantized string. 

Now, let us pay close attention to the $\kappa_I < 2\kappa$ case. 
As it can be easily verified, $\dot{A}$ is finite in this case. 
This implies that each $|B_n|^2$ is finite. First, we shall assume that 
the solution of Eq.~(\ref{eq-field}) can be represented perturbatively 
such as, 
\begin{eqnarray}
\label{eq-perturv}
x_n = e^{-in \tau} + \delta x_n(\tau), 
\end{eqnarray}
where we start with a purely positive frequency solution 
$x_n = e^{-in \tau}$ for $\tau\le -\tau_0$. 
Substituting Eq.~(\ref{eq-perturv}) into 
Eq.~(\ref{eq-field}), we find 
\begin{eqnarray}
\label{eq-deltaX}
\ddot{\delta{x_n}}+n^2 \delta x_n \simeq H(P\tau)P^2 e^{-in\tau}. 
\end{eqnarray}
By multiplying the above equation by $e^{-in\tau}$ and integrating 
from $\tau=-\tau_0$ to zero, we can get the following relation, 
\begin{eqnarray}
\label{eq-deltaX1}
0
\dot{\delta{x_n}}(0)+i n \delta x_n(0) \simeq 
\int_{-\tau_0}^0 H(P\tau)P^2 e^{-2in\tau} d\tau. 
\end{eqnarray}
Remind that $x_n\simeq e^{-in\tau}+ B_n e^{in\tau}$ in $\tau\ge 0$. Then, 
\begin{eqnarray}
\label{eq-deltaX2}
2in B_n &\simeq& 
\int_{-\tau_0}^0 H(P\tau)P^2 e^{-2in\tau} d\tau \nonumber \\
 &\sim& -\frac{\alpha P}{{r_I}^2}\left(\frac{P}{2n} \right)^{1-\gamma}
\int_{0}^{2n \tau_0} t^{-\gamma} e^{-it} dt, 
\end{eqnarray}
where $\gamma$ is defined as 
\begin{eqnarray}
\label{eq-gammadef}
\gamma:=-\frac{2(\kappa-\kappa_I)}{\kappa_I}. 
\end{eqnarray}
This calculation is essentially the same as in~\cite{VegaSanchez}. 

To see whether $\langle N_{\rm{out}} \rangle$ and 
$\langle {M^2}_{\rm{out}} \rangle$ diverge, 
we have only to calculate $|B_n|^2$ for the high frequency modes, $n \gg 1$. 
Thus, the values of $|B_n|^2$ for the high frequency modes are as follows, 
\begin{eqnarray}
\label{eq-Bnhigh}
B_n &\sim& \frac{i\alpha }{{r_I}^2}\left(\frac{P}{2n} \right)^{2-\gamma}
\int_{0}^\infty t^{-\gamma} e^{-it} dt \nonumber \\
 &=& i^{2-\gamma}\frac{\alpha}{{r_I}^2}\left(\frac{P}{2n} \right)^{2-\gamma} 
\Gamma(1-\gamma), 
\end{eqnarray}
where $\Gamma(x)$ is the gamma function. Because $0<\gamma<1$ 
by the definition of 
$\gamma$~(\ref{eq-gammadef}), $|B_n|^2$ is well defined and we obtain 
immediately 
\begin{eqnarray}
\label{eq-N-Mn}
\langle N_{\rm{out}} \rangle &\propto& \sum_n n^{-4+2\gamma}\sim 
\mbox{finite}, \nonumber \\
\langle {M^{2}}_{\rm{out}} \rangle &\propto& \sum_n n^{-3+2\gamma}\sim 
\mbox{finite}. 
\end{eqnarray}
This strongly suggests that a first-quantized string can cross the CH 
without infinite excitation and that the horizon can be singularity-free in 
terms of the quantized string. 

\begin{figure}[htbp]
 \centerline{\epsfxsize=9.0cm \epsfbox{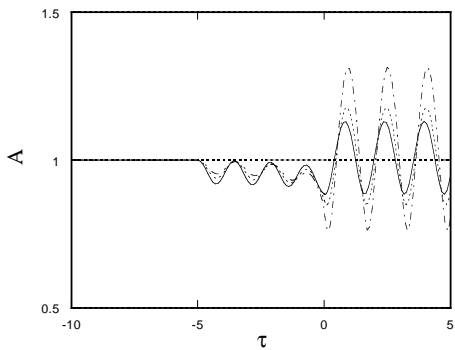}}
         \caption{$\tau$-$A$ diagram of the $n=2$ mode for the $0<\gamma<1$ 
case. We set $\alpha/r_I=1$ and $\gamma=0.2$ (dot-dashed line),\,
$0.4$ (dashed line),\,$0.6$ (solid line). The ingoing flux is turned 
on at $\tau=-\tau_0=-5$. Both $A$ and $\dot{A}$ are finite even 
on the CH~($\tau=0$). Hence we can define the Bogoliubov coefficients 
in the $\tau \to 0$ region.}
               \protect
\label{fig-2}
\end{figure}

\begin{figure}[htbp]
 \centerline{\epsfxsize=9.0cm \epsfbox{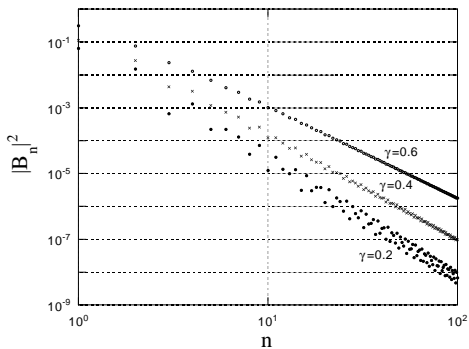}}
         \caption{The Bogoliubov coefficient for each $n$ mode and 
different parameters $\gamma$. We can see that the Bogoliubov 
coefficients decrease faster than $\sim n^{-2}$ for $0<\gamma<1$.}
               \protect
\label{fig-3}
\end{figure}

Now we shall confirm numerically that our perturbative treatment is 
correct. We solve the equation~(\ref{eq-Adiff}) with the 
potential~(\ref{eq-H2}). 
Here we set the amplitude of the potential $\alpha P^2/r_I^2=1$ 
and $A=1$ at $\tau=-\tau_0$. The ingoing flux is turned on at $\tau_0 = 5$. 
Even for different values of these parameters, the results are qualitatively 
the same. 
We show the amplitude $A$ for $n=2$ and $\gamma = 0.2$, $0.4$,
$0.6$ in Fig.~\ref{fig-2}. We see that after the ingoing flux is turned on 
($\tau \ge- \tau_0$), $A$ oscillates and the test string is excited by 
the ingoing flux. Both $A$ and $\dot{A}$ stay finite and continuous 
even at the CH. 
Hence we can define the Bogoliubov coefficients in the $\tau>0$ 
region and calculate the expectation value of the mass-squared operator. 
Fig.~\ref{fig-3} shows the $n$-dependence of the Bogoliubov coefficients for 
each value of $\gamma$. Since they decrease faster than $n^{-2}$, the 
expectation value of the mass-squared operator converges. 

\begin{figure}[htbp]
 \centerline{\epsfxsize=9.0cm \epsfbox{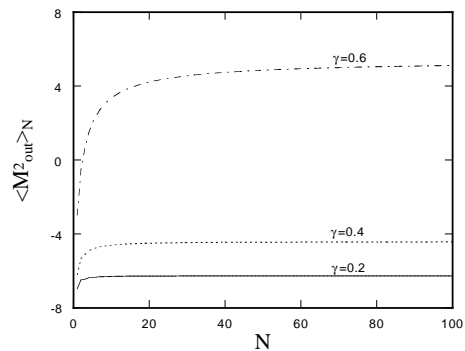}}
         \caption{The expectation value of the mass-squared operator 
for different parameters $\gamma$. We can see that 
${\langle {M^2}_{\rm{out}} \rangle}_N$ approaches a constant value 
in the $N \to \infty$ limit, which means that the expectation 
value of the mass-squared operator is finite for $0<\gamma<1$. Hence 
the test string can pass across the CH.}
               \protect
\label{fig-4}
\end{figure}

Finally we show the expectation value of the mass-squared operator in 
Fig.~\ref{fig-4}. 
${\langle {M^2}_{\rm{out}} \rangle}_N$ means that the sum is taken up to $N$. 
Hence the real expectation value is realized in the $N \to \infty$ limit. 
As we expected, ${\langle {M^2}_{\rm{out}} \rangle}_N$ converges for each 
$\gamma$, which means that the excitation of the string is finite. 
Hence the string can smoothly pass through the CH for $0<\gamma<1$, i.e. 
$\kappa_I< 2\kappa$. 

\section{Concluding remarks}
We have examined whether a null weak singularity inside generic 
charged black holes is a real end of spacetime in terms of a first-quantized 
test string by using a charged Vaidya model. In an asymptotically flat 
spacetime~($\Lambda=0$ case), 
the first-quantized test string falling into the CH is always excited 
infinitely, while it is not when $\kappa_I<2\kappa$ in an asymptotically 
de Sitter spacetime~($\Lambda> 0$ case), in spite of the existence of 
a p.p. curvature singularity along the CH. 
We should note that the latter case occurs in the physically relevant 
range of parameters $m>q$. 
This implies that the asymptotically de Sitter spacetime can be extended 
through the CH and hence that the strong cosmic censorship is 
violated in string theory. 

It is worth commenting that when the string is excited infinitely, 
one should further consider for the following two points: (i) in 
general, a second-quantized string theory is necessary for judging whether 
the CH is the real end of spacetime in string theory; (ii) the plane-wave 
approximation is violated in the solutions near the CH. As for the first 
point, 
however, Horowitz and Steif~\cite{HS2} speculated that there is no ``string 
creation'' in the plane-wave metric from the fact that 
there is no particle creation in the metric~\cite{Gibbons}. This suggests 
that the first-quantized description should be adequate in the spacetime 
under consideration. As for the second point, the typical size of the 
string grows to infinity as follows~\cite{MitchellTurok}, 
\begin{eqnarray}
\label{eq-size}
<r^2>\,\sim \int d\sigma :x^i(0,\sigma)^2:\, \sim \sum_{i, n} 
\frac{|B_n^i|^2}{n}\sim \infty, 
\end{eqnarray}
where $:\,\,:$ denotes the normal ordering required to make 
the quantum operator well defined. Therefore, it is still an open 
question how much this violation would affect our results in the 
asymptotically flat case. 

As already mentioned before, the internal structure of generic Kerr-type 
vacuum black holes is locally quite similar to that of spherically 
symmetric charged black holes. 
Thus, as verified in the charged Vaidya model, if the surface gravity of 
the inner horizon in the generic Kerr-type black holes is small enough, 
we naively expect that the spacetime near the CH is approximately described 
by the plane-wave metric, 
\begin{eqnarray}
\label{eq-plane-waveK}
ds^2 \approx -du\,dv + F(v,x^i)dv^2 +dx_i dx^i, 
\end{eqnarray}
\begin{eqnarray}
\label{eq-plane-waveKF}
F = W_{ij}(v) x^i x^j, \nonumber 
\end{eqnarray}
where $W_{ij}(v)$ is symmetric and traceless from the vacuum Einstein 
equations~(here we suppose that the cosmological constant is negligibly 
small). In this case, the metric is also a metric of the classical solution 
for string theory~\cite{HS1} and the same result as in the charged 
Vaidya model would be obtained. 

In general, the back reaction from the first-quantized string 
and the quantum corrections for the background metric should be considered. 
However, such effects should be very small in the case that a first-quantized 
test string is not excited infinitely. This strongly suggests that the strong 
cosmic censorship is violated in an asymptotically de Sitter spacetime in 
string theory, in contrast to general relativity. 

\section*{acknowledgment} 
This paper owes much to the thoughtful and helpful comments of Shigeaki 
Yahikozawa. We would like to thank Akio Hosoya, Hideo Kodama, Daniel 
Sudarsky, and Akira Tomimatsu for useful discussions. 
Our special thanks are due to Albert Carlini for reading our manuscript 
carefully. K.~M. is also grateful to Kei-ichi Maeda for providing 
me with continuous encouragement. 
This work is supported in part by Scientific Research Fund of the 
Ministry of Education, Science, Sports, and Culture, by the 
Grant-in-Aid for JSPS~(No. 199704162 and No. 199906147).

\end{document}